\documentclass[nobm]{epl}

\newcommand{\V}{V^{\ast}}
\newcommand{\e}{\varepsilon}  
\newcommand{\eq}[1]{eq.~(\ref{#1})}               
\graphicspath{{Figures/}}

\title{``Marginal pinching'' in soap films}
\author{A.\ Aradian \thanks{E-mail \email{Achod.Aradian@college-de-france.fr}} 
\and E.\ Rapha\"el \and P.-G.\ de Gennes}
\institute{Physique de la Mati\`ere Condens\'ee, 
CNRS URA 792, Coll\`ege de France - \\ 11, place Marcelin Berthelot, 75231 Paris Cedex 05, France.\\
}
\pacs{68.15.+e}{Liquid thin films}
\pacs{83.80.Iz}{Emulsions and foams}
\pacs{02.40.Xx}{Singularity theory}

\begin{document}

\maketitle

\begin{abstract}
We discuss the behaviour of a thin soap film facing a frame 
element: the pressure in the Plateau border around the frame is 
lower than the film pressure, and the film thins out over a 
certain distance $\lambda (t)$, due to the formation of a 
well-localized pinched region of thickness $h(t)$ and extension 
$w(t)$. We construct a hydrodynamic theory for this thinning 
process, assuming a constant surface tension: Marangoni 
effects are probably important only at late stages, 
where instabilitites set in. We find $\lambda (t) \sim t^{1/4}$, 
and for the pinch dimensions, $h(t) \sim t^{-1/2}$ and $w(t) 
\sim t^{-1/4}$. These results may play a useful role 
for the discussion of later instabilitites 
leading to a global film thinning and drainage, as first 
discussed by K.\ Mysels under the name ``marginal regeneration''.
\end{abstract}
Early experiments by K.\ Mysels and coworkers~\cite{FrankelMyselsLivre}
showed that a vertical soap film, suspended on a frame, (and made 
with ``mobile'' surfactant) thins out by nucleation and growth of 
black, thin spots near the Plateau borders. They called this process 
``marginal regeneration''.

There are in fact (at least) two steps in marginal regeneration:
a) The pressure in the Plateau border is lower than the 
pressure in the film. This thins out the film near the border, and 
leads to a ``pinch''.
b) The pinched state must have an intrinsic instability 
leading to the black spots.
The two steps are very different: the pinch can occur at constant 
surface tension---i.e.\ without any Marangoni effect. On the other 
hand, the later instabilities are probably triggered by Marangoni 
flows, as pointed out by a number of authors~\cite{JoyeInstability,
BruinsmaRegeneration,SteinRegeneration,NierstraszFrensRegeneration}.

Our aim in the present note is restricted to the first step,
i.e. the description of the pinched state with its dynamics.
Pinching has already been studied in connection with 
the elimination of dimples in the coalescence of drops~\cite{
FrankelMyselsCoalescence,JonesWilsonDrainage} or in 
the drainage of thin films~\cite{JoyeAxisymmetrical,
BertozziSingularities}. However, in these problems, 
the extension of the dimpled zone is prescribed and fixes one of the 
spatial scales involved. Our case is different: we consider a 
semi-infinite film (of initial thickness $e_0$) facing a straight 
Plateau border. At $t=0$ the film begins to pinch, and is 
perturbed over a certain distance $\lambda (t)$ increasing with 
time (as we shall see $\lambda (t) \sim t^{1/4}$). Also the width 
$w(T)$ of the small pinched region decreases with time ($w(t) \sim 
t^{-1/4}$) and so does the thickness of the pinch $h(t) \sim 
t^{-1/2}$. It seems important to know these scales in detail 
before embarking into the second step: all data on dimples 
suggest that instabilities can only occur after a long stage of 
pinching. This is why we study the pinch in the simplest condition 
of a constant surface tension. In this case, as will be shown in this 
article, a complete, explicit solution involving the the two scales 
$\lambda$ and $w$ can be constructed.

\begin{figure}
\onefigure[scale=0.9, clip=true, bb=3.2cm 18.5cm 17cm 24cm]{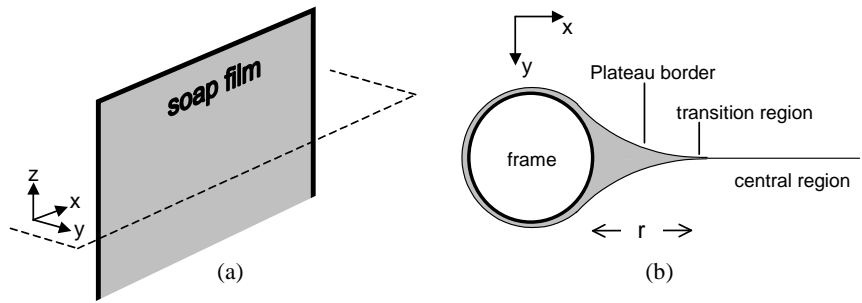}
\caption{(a) Soap film held in a vertical frame, and $xy$ cross-section plane. 
(b) Sketch of the initial shape in the cross-section plane (here, a case 
where the liquid totally wets the frame material). As the film in the central 
region is very thin (uniform half-thickness $e_0 \ll r$), it has been represented 
simply as a line.}
\label{filmview}
\end{figure}

We thus consider a soap film, held in a vertical rectangular frame as in 
fig.~\ref{filmview}-a, and we wish to describe the time evolution of 
the film profile in a horizontal cross-section, neglecting in the process
all drainage in the vertical direction (vertical drainage in ``mobile''
films is tightly associated with the second step of marginal regeneration, 
and will intervene significantly only once the instabilities have risen up). 
The cross-section profile is therefore treated as a one-dimensional thin film 
of half-thickness $e(x,t)$ at position $x$ and time $t$.

The initial shape of the film has the following general 
features~\cite{FrankelMyselsLivre} (see fig.~\ref{filmview}-b): 
joining the frame legs is the ``Plateau border'' of size $r$, where 
the thickness is of macroscopic dimensions. In this region, 
the curvature of the film is prescribed to a fixed value $c$. 
Further away from the frame legs, there is a transition region, 
where the thickness decreases steeply. And, finally, we reach the 
zero-curvature central region, of constant thickness $2 e_0$,
which extends over the major part of the frame width.
The system displays in addition a natural separation of spatial scales. 
Firstly, the width of the frame (e.g., $10 \,$cm) is supposed to be much 
larger than the Plateau border extension $r$ (e.g., $1 \,$mm).
We shall thus assume that the width of the frame can safely be taken as 
infnite in regard of the length $r$, leaving us with a 
semi-infinite film. Secondly, the Plateau extension $r$ itself is much 
larger than the limit thickness $e_0$ in the central region 
(e.g., $10 \, \mu$m). Our study of the film dynamics will therefore 
rely on an asymptotic approach~\footnote{This kind of approach has been initiated and 
developed by Jones and Wilson~\cite{JonesWilsonDrainage} in the resolution of a 
closely related coalescence problem.} valid in the limit $\e \ll 1$, where 
$\e$ represents the aspect ratio of the film: 
\begin{equation}
\label{epsilondef}
\e = e_0/r \, .
\end{equation}

The evolution equations are classically written in the lubrication 
approximation. The axis are shown on fig.~\ref{filmview}-b, and, for 
simplicity, we choose the origin of the $x$-axis at the position 
of minimum thickness, in the pinch~\footnote{We assume that, 
once the pinch is formed, its spatial position does not 
change anymore, or, at least, that its motion remains of subdominant 
importance to the solution. This hypothesis is indeed supported by 
previous work on dimple formation in analogous situations~\cite{FrankelMyselsCoalescence,
JoyeAxisymmetrical,BertozziSingularities}.}. Assuming that velocity
of the liquid vanishes at the film/surfactant layer interface, 
the flow is of Poiseuille type and yields a parabolic velocity profile. With a 
Laplace-Young pressure $p=p_{\ab{air}}-\gamma e_{xx}$ in the fluid, we 
find that the flux $Q$ in the $x$-direction is given by $Q=\frac{2}{3} \V 
e^3e_{xxx} \, $. Here, $\V=\gamma/\eta$ is a characteristic capillary 
velocity and is defined as the ratio of the surface 
tension $\gamma$ by the viscosity of the liquid $\eta$. ($\gamma$ might 
depend on the vertical elevation of the considered cross-section, but 
is otherwise constant in the $xy$-plane.) 
Inserting into the continuity relation $2 e_t + Q_x = 0 \, $, we find the 
governing equation of the film profile:
\begin{equation}
\label{governing}
e_t + \frac{\V}{3}(e^3e_{xxx})_x=0 \, .
\end{equation}
Note that microscopic interactions~\cite{BergeronReview} like van der Waals, 
or electrical interactions, have not been included in \eq{governing}, 
since we are mainly interested in thicknesses larger than their range of
action. We will simply consider that once the film thickness reaches 
small enough values ($ \simeq 100 \,$\AA), these additional interactions 
are globally repulsive and prevent any further thinning.

It is more convenient to use the reduced variables (denoted with capital letters) 
$X=x/r \,$, $E=e/e_0 \,$, and $T=t/t_{\ab{relax}} \,$, where 
$t_{\ab{relax}}=\frac{3}{\V} r^4/e_0^{\ 3}$ is a typical relaxation 
time needed for a bump of height $e_0$ and width $r$ to be leveled off
by capillary flow~\footnote{ Here is a simple way to see this: the capillary 
flux, as used in establishing \eq{governing}, is $Q \simeq \V 
e^3e_{xxx} \sim \V e_0^{\ 4}/r^3 \,$. Then, $t_{\ab{relax}}$ 
appears as the time needed to empty the volume of the bump 
with such a flux, i.e., $t_{\ab{relax}} \simeq e_0 r/Q \,$.}. 
In these reduced variables, \eq{governing} takes the dimensionless form
\begin{equation}
\label{dimensionless}
E_T + (E^3E_{XXX})_X=0 \, .
\end{equation}

From the initial shape described above, a pinch has first to form. Describing this 
process requires the precise knowledge of the initial state. However, 
it is expected that the formation stage is quite rapid on the time 
scale of $t_{\ab{relax}}$, and, moreover, the details of it appear 
in fact to be unessential to the longer term dynamics of the film.
We may nevertheless understand why only a localized region thins out 
at the beginning and starts to pinch: within the initial profile, 
the pressure gradient is non-vanishing only over a very restricted 
zone, inside the transition region (there, the curvature decreases 
from the border value $c$ to the central value zero).

We will hereafter simply assume that this early stage leads to the formation of 
a pinch precursor of reduced thickness. Without the need of further knowledge, 
we are then in a position to describe the evolution 
of the soap film. In order to do so, we solve the equation governing 
the dynamics (\eq{dimensionless}) as follows: we divide the film into three 
spatial regions (inherited from the three regions in the initial profile),  
and look for solutions in each of them, imposing in addition that these 
solutions all match together (in the sense of symptotic matching). 
Starting from the frame leg, the three regions are: the Plateau border, 
(at negative values of $X$), then the pinch region that lies in the vicinity 
of $X=0$, and, eventually, the central region, which extends 
over the positive values of $X$. 

The profile in the Plateau border is easily determined to 
leading order in the asymptotic parameter $\e$: as the border contains 
a great mass of liquid compared to the rest of the film, it remains 
unchanged during the whole evolution. In particular, 
its curvature remains the same as initially (i.e., $e_{xx}=c \,$), maintaining
thereby a constant pressure drop as compared to the pressure in the central region. 
This feature will naturally prove essential to the film dynamics.

For clarity's sake, it turns out being more convenient to pursue 
with the description of the central region ($X>0$). The existence 
of a pinch at $X=0$, with a 
sudden thickness decrease, involves that the central region, 
initially flat, is pushed out of equilibrium, and that a flow 
must take place in order to relax, to the greatest possible 
extent, the curvature induced by such pinching. But this flow
cannot easily pour liquid through the pinch region, whose small dimensions 
strongly reinforce viscous resistance. Thus, the relaxation
flow is mostly directed towards the far-field zone ``at 
infinity'' ($X \gg 1$). In a more formal way, the relaxation 
flow $Q_{\ab{relax}}$ towards the far-field zone is of order unity, 
whereas the flow poured through the pinch $Q_{\ab{pinch}}$ 
is only of order $\e$ (as will be shown later on). 
Thus, in the spirit of boundary-layer techniques, we look inside 
the central region for a solution at zeroth-order in $\e$ 
(a so-called ``outer'' solution), that is to say that the relaxation 
process is solved with a boundary condition of \emph{zero} flux 
at $X \ll 1$ (in the vicinity of the pinch region). The complete set
of boundary conditions for the central region solution writes:
\begin{equation}
\label{boundarycentral}
\mbox{(i)}\quad \left. E \right|_{X \ll 1}=0 \, , \qquad 
\mbox{(ii)}\quad \left. E^3E_{XXX} \right|_{X \ll 1}=0 \, , \qquad 
\mbox{(iii)}\quad \left. E \right|_{X \gg 1} \to 1 \,.
\end{equation}
The first condition imposes that the film thickness must vanish
(at zeroth-order in $\e$) at the approach of the pinch, the 
second one is the zero flux condition, and the third one ensures 
that the far-field film remains unperturbed and 
retains the initial thickness $e_0$.

\begin{figure}
\twoimages[clip=true, bb=3cm 7cm 18cm 20.3cm, width=6.9cm]{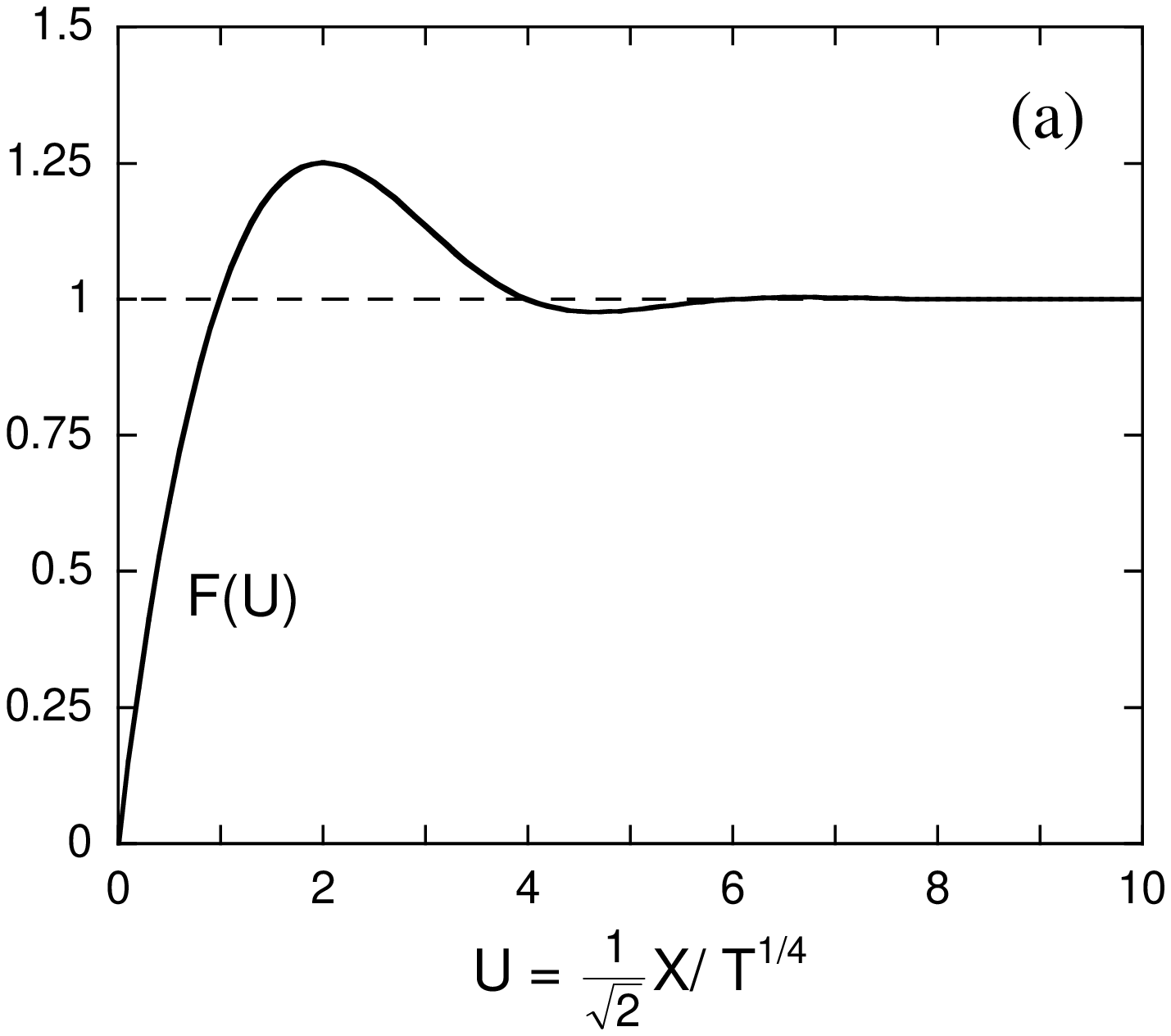}{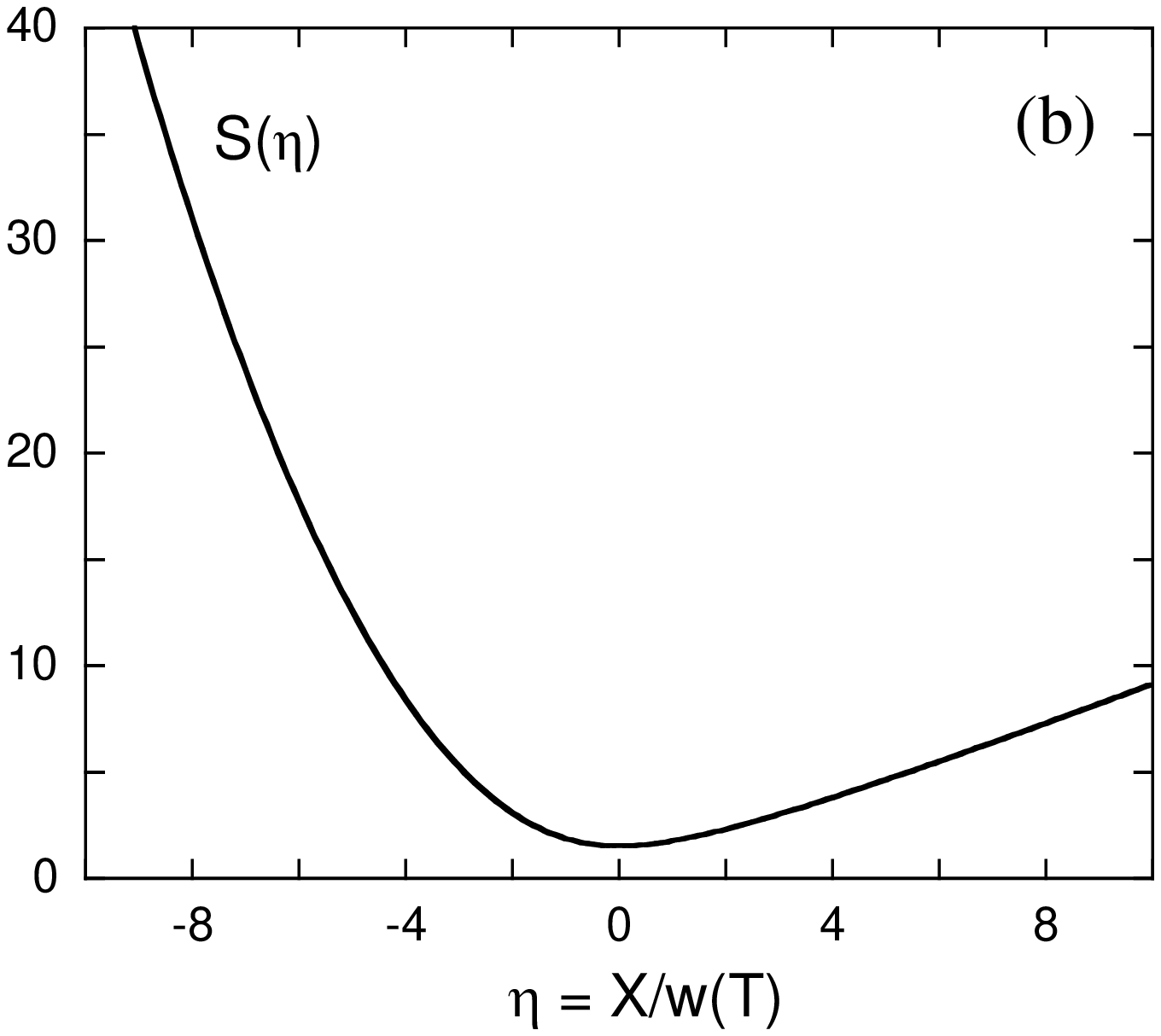}
\caption{(a) The self-similar function $F(U)$ describing the film profile in the central region.
(b) The self-similar function $S(\eta)$ describing the pinch.}
\label{FandSplot}
\end{figure}

We propose a solution to eqs.~(\ref{dimensionless}) 
and~(\ref{boundarycentral}) in a self-similar form, with the 
self-similar variable $U=\frac{1}{\sqrt{2}}X/T^{1/4}$. 
Setting $E(X,T)=F(U)$, and inserting into
\eq{dimensionless}, we obtain the following equation for 
the self-similar function $F$:
\begin{equation}
\label{Fequation}
U F'=(F^3 F''')' \, .
\end{equation}
A study of the film profile in the pinch region (see further down)
shows that a matching with the central region is possible if 
the behaviour of $F$ near the origin is either linear, parabolic, 
or power-law (with exponent 3/4). However, it turns out that only the 
linear behaviour is compatible with eqs.~(\ref{boundarycentral}) 
and~(\ref{Fequation}), i.e., $F \sim A U$ for $U \ll 1$, with $A$ a constant. 
The complete function $F$ can be constructed numerically with the help
of a two-sided shooting procedure based on analytical 
expansions at $U \ll 1$ and $U \gg 1$ (we omit technical details 
for conciseness). The resulting plot is given on fig.~\ref{FandSplot}-a: 
as one can see, $F$ increases from zero to unity over a range of a few 
units in the variable $U$, and then presents decreasing capillary oscillations. 
The shooting procedure also determines the numerical value of the slope of 
F at the origin: $A \simeq 1.591$.

The important physical consequence of these results is that, because of 
the presence of the pinch at $X=0$, the film in the central region is 
also thinned out, and retrieves the far-field thickness $e_0$ over a distance 
$\lambda(t) \simeq r T^{1/4} \sim t^{1/4}$, which increases with time.                        
 
We are thus in possession of a self-similar solution of the problem 
in the central region. One may though wonder about the 
relevance of self-similarity in describing the real physical 
solution. Actually, if we take a 
version of \eq{dimensionless} that is linearized around $E=1$, and solve 
it analytically with the Green function method with boundary 
conditions similar to \eq{boundarycentral}, we rigorously
find that the solution indeed takes the above self-similar form. 
We conjecture here that this feature can be safely carried over to 
the non-linear case.
                                        
We next consider the solution in the pinch region. As stated earlier, 
once the pinch is well-formed, the suction from the Plateau border is 
able to drain liquid through the pinch out of the central region,
but only with an asymptotically small current $Q_{\ab{pinch}}$.
However, the dimensions of the pinch are so small, that we can reasonably
take this current to be spatially uniform over it~\cite{FrankelMyselsCoalescence,
JonesWilsonDrainage,BertozziSingularities} (but not constant in 
time): $Q_{\ab{pinch}}=Q_{\ab{pinch}} (T)$. Therefore, to leading order, 
the profile of the film in the pinch 
region obeys the following uniform current equation, 
rather than the full \eq{dimensionless}:
\begin{equation}
\label{pinchequation}
E^3E_{XXX}= Q_{\ab{pinch}}(T) \, .
\end{equation}
Here again, we try a self-similar solution of the form 
$E(X,T)=h(T)S(\eta)\,$, with $\eta=X/w(T)\,$. Such a solution has 
been proposed in references~\cite{JonesWilsonDrainage,
BertozziSingularities}. In this solution, the functions $h(T)$ and $w(T)$ 
represent, respectively, the typical \emph{height} and \emph{width} of the pinch, 
and will be determined by matching with the neighbouring Plateau and
central regions. Replacing the self-similar form in 
\eq{pinchequation}, we get $S^3 S''' \: h(T)^4/w(T)^3=
Q_{\ab{pinch}}(T)\,$, from what we deduce that $S^3S'''$ 
cannot bear any dependence on $\eta$ (this would bring in the l.h.s 
of the equation a dependence on $X$ that could not be balanced by the 
r.h.s.). Thus, the differential equation on $S$ 
is necessarily
\begin{equation}
\label{Sequation}
S^3S'''=-\alpha \, ,  
\end{equation}
where $\alpha$ is a positive constant (the sign has been chosen 
in view that $Q_{\ab{pinch}}\,$, oriented towards the
negative coordinates, is negative). At this point, $\alpha$ is still unknown, 
but shall be determined later on. 

Let us now apply the matching requirements that the self-similar 
solution must fulfill, starting by the match with the Plateau border. 
A study of the asymptotic behaviours permitted to $S$ by \eq{Sequation}
for large $\eta$ (either positive or negative), opens three possibilities, 
namely linear, parabolic or power-law with exponent 3/4. The Plateau border 
imposes a constant curvature $e_{xx}=c \,$, or in terms of the reduced variables, 
$\left. E_{XX} \right|^{\ab{Plateau}}_{X \to 0} = rc/\e \,$. Thus, on the pinch side,
$S$ must  be asymptotically parabolic. Equating the Plateau curvature with
the pinch curvature $\left. E_{XX} \right|^{\ab{pinch}}_{\eta \to -\infty} = \lim 
S'' \: h(T)/w(T)^2$ yields moreover the two conditions:
\begin{equation}
\label{matchingpinch}
\mbox{(i)} \quad h(T)/w(T)^2=rc/\e \, , \qquad 
\mbox{(ii)} \quad \lim_{\eta \to -\infty} S''(\eta)=1.                  
\end{equation}
We proceed further and match now with the central region. 
As stated above, there are three possibilities of 
asymptotic behaviour of $S$ at the approach of the 
central region, but on the central region side, we found 
that only the linear one is consistent for the function $F$ 
with the boundary conditions~(\ref{boundarycentral} and the differential 
equation~(\ref{boundarycentral}), giving $\left. E(X,T)\right|_{X \ll 1}^{\ab{central}} 
\simeq A U =\frac{A}{\sqrt 2} X/T^{1/4} \,$. In the pinch, $S(\eta)$ must accordingly
have an asymptotically linear behaviour (for $\eta \gg 1$), which leads to the 
corresponding expression of $E$: $\left. E(X,T) \right|_{\eta \gg 1}^{\ab{pinch}}
= h(T) \, \eta \, \lim S' = X \, h(T)/w(T) \, \lim S'$. Matching these central and pinch 
behaviours together, we find that two new conditions must be met:
\begin{equation}
\label{matchingpinch2}
\mbox{(i)} \quad h(T)/w(T)=\frac{A}{\sqrt{2}} T^{-1/4}\, , \qquad 
\mbox{(ii)} \quad \lim_{\eta \to \infty} S'(\eta)=1 \, .
\end{equation}

With eqs.~(\ref{matchingpinch}-i) and~(\ref{matchingpinch2}-i), 
we are now able to extract the expressions of $h$ and $w$:
\begin{equation}
\label{handw}
h(T)=\e \: \frac{A^2}{2rc} T^{-1/2} \, , \qquad 
w(T)=\e \: \frac{A}{\sqrt2 rc} T^{-1/4} \, .                 
\end{equation}
We immediately notice that, as announced before, $h$ and $w$, the 
typical dimensions of the pinch, are of order $\e$ and hence 
asymptotically smaller than the dimensions (of order unity) that prevail 
in the central region. Furthermore, we see that, as time passes, the pinch becomes 
thinner and smaller in extent. We may also note that the time dependences in 
\eq{handw} are, not too surprisingly, equal to the ones found in closely 
related drainage problems~\cite{FrankelMyselsCoalescence,JonesWilsonDrainage,
BertozziSingularities}.

To complete our solution of the profile in the pinch, we still 
have to exhibit a function $S$ that has the correct shape, as 
dictated by eqs.~(\ref{Sequation}), (\ref{matchingpinch}-ii) 
and~(\ref{matchingpinch2}-ii). Such a function has indeed been 
computed previously~\cite{FrankelMyselsCoalescence,
JonesWilsonDrainage}. At large $\eta$, $S$ admits an expansion 
around the imposed linear profile of slope 
unity, from which we have started numerical integrations 
towards negative values of $\eta$. The integrations have shown that 
complying with the requirement that $S''$ equals unity for large negative 
$\eta$ ascribes to the parameter $\alpha$ in \eq{Sequation} the value 
$\alpha \simeq 1.21 \,$. The minimum value of $S$ (at $\eta=0$) is then 
found to be $S_{\ab{min}} \simeq 1.52 \,$. A plot of $S$ is provided in 
fig.~\ref{FandSplot}-b.

With the knowledge of the self-similar functions $F$ and $S$, 
we have now at disposal a complete description of the profile 
of the soap film, once the pinch is formed (and under the basic hypothesis 
that it remains later on fixed in position). We may at this point give 
justifications to some assumptions that were made in the course of the resolution.

First, it was assumed that to zeroth order in $\e$, one was entitled 
to solve the central region profile with zero flux towards the 
pinch (eq.~(\ref{boundarycentral}-ii)). This can be justified as follows.
In the central region, the current generated by the relaxation process is, 
by the very definition of the reduced variables, of order one: 
$Q_{\ab{relax}}=\mathcal{O}(1)$. On the other hand, the expression of the current 
in the pinch was given when establishing \eq{Sequation}: $Q_{\ab{pinch}}
=S^3 S''' \: h(T)^4/w(T)^3 \,$. Since both $h$ and $w$ are of order $\e$, we 
deduce that $Q_{\ab{pinch}}=\mathcal{O}(\e) \ll Q_{\ab{relax}} \,$, and is
thus negligible to leading order in the central region.

A second approximation 
was that, since the pinch was of small dimensions, the current $Q_{\ab{pinch}}$ 
could be taken uniform inside it. Indeed, using the continuity equation, 
one can estimate the thereby neglected spatial variation $\delta Q_{\ab{pinch}}$ 
as the variation in time of the volume enclosed in the pinch region: we 
have $\delta Q_{\ab{pinch}} \simeq \frac{\upd}{\upd t}(wh) = \mathcal{O}(\e^2) \ll  
Q_{\ab{pinch}} \,$, as had been assumed.

In addition, it has been assumed, when establishing \eq{governing}, 
that the velocity field goes to zero at the film/surfactant 
layer interface, but also that the surface tension $\gamma$ keeps a 
constant value throughout the film. This cannot be strictly true: For the
surfactant layer to be at rest, it must develop a surface tension gradient to 
counter the friction force generated by the underneath liquid flow. We can estimate 
the required gradient by equating it with the viscous stress at the surface:
$(1/\gamma_0) \partial \gamma/\partial x = \left. (\eta/\gamma_0) \partial v
/\partial y \right|_{y=e} = e e_{xxx}$ (the equations were divided through
by $\gamma_0$, the constant tension that would be obtained in the absence 
of liquid flow). Using this formula with our solution for $e(x,t) \,$, we find that the 
relative variation of the surface tension across the pinch is 
$\left. \delta \gamma/\gamma_0 \right|_{\ab{pinch}} = \int_{\ab{pinch}} 
(1/\gamma_0) \partial \gamma/\partial x \, \upd x \simeq w (e_0^{\ 2}/r^2)(h^2/w^3) 
= \mathcal{O}(\e^2) \,$. Similarly, we find that the variation over the characteristic 
length $r$ in the central region writes $\left. \delta \gamma/\gamma_0 
\right|_{\ab{central}}= \mathcal{O}(\e^2) \,$. Thus, the gradients required to cancel 
the flow velovity are indeed very small, and the surface tension can safely 
be taken as constant (at least, in the study of the leading-order dynamics of 
the film). However, we should point out that this result is by no means 
contradictory with the later destabilization by Marangoni flows, as these 
occur in the \emph{vertical} direction.  
    
A final remark is in order concerning the late stage of the film 
pinching. As mentioned earlier, we expect that, at small scales, 
repulsive microscopic forces stop the thinning (with, 
depending on the electrolyte concentration in the film,
different possible scenarii~\cite{JoyeAxisymmetrical}). The corresponding 
time $t_{\ab{stop}}$ can be roughly estimated, if we say that the stop 
occurs for $e$ around $100 \,$\AA, and use \eq{handw} with typical 
numerical values (for instance, $e_0 \simeq 10 \, \mu$m, 
$r \simeq 1/c \simeq 1 \,$mm, $\V \simeq 10^2 \,$m/s): we find 
$t_{\ab{stop}} \simeq 10^3 \,$s. We should caution that, in regard with 
experiments on marginal regeneration, this is a significative duration, 
for which one would then probably have to include the study of the second 
step of marginal regeneration (Marangoni instabilities). The thinning process might 
therefore not be allowed to proceed by itself as far as $t_{\ab{stop}}$. 
We can also estimate the distance over which the central region has 
a thickness departing from $e_0$ (assuming that the thinning keeps on 
until $t_{\ab{stop}}$): $\lambda_{\ab{stop}} \simeq r (t_{\ab{stop}}/t_{\ab{relax}})^{1/4} 
\simeq 3r \simeq 3 \,$mm. This distance is much smaller than the frame width, 
justifying our description of the film as semi-infinite towards the center 
of the frame. However, in situations where the frame width is not so large, our 
central region solution should be regarded as transient: when $\lambda (t)$ 
becomes of the order of the frame width, we expect to retrieve the solutions 
in bounded geometries that have been described by several authors~\cite{
FrankelMyselsCoalescence,JonesWilsonDrainage,BertozziSingularities} 
(for instance, a quasi-static parabola in the central region).

\acknowledgments
We are pleased to thank H.\ A.\ Stone for very valuable help and A.\ L.\ Bertozzi 
for instructive discussions.


\begin{thebibliography}{0}



\bibitem{FrankelMyselsLivre}
  \Name{Mysels K. J., Shinoda K. \and Frankel S.}
  \Book{Soap Films, Studies of Their Thinning and a Bibliography}
  \Publ{Pergamon Press, New York}
  \Year{1959}.
  
\bibitem{JoyeInstability}
  \Name{Joye J.-L., Hirasaki G. J. \and Miller C. A.}
  \REVIEW{Langmuir}{10}{1994}{3174};
  \REVIEW{J. Colloid Interface Sci.}{177}{1996}{542}.       

\bibitem{BruinsmaRegeneration}
  \Name{Bruinsma R.}
  \REVIEW{Physica A}{216}{1995}{59}.

\bibitem{SteinRegeneration}
  \Name{Stein H. N.}
  \REVIEW{Adv. Colloid Interface Sci.}{34}{1991}{175}.

\bibitem{NierstraszFrensRegeneration}
  \Name{Nierstrasz V. A.\and Frens G.}
  \REVIEW{J. Colloid Interface Sci.}{207}{1998}{209}.

\bibitem{FrankelMyselsCoalescence}
  \Name{Frankel S. P. \and Mysels K. J.}
  \REVIEW{J. Phys. Chem.}{66}{1962}{190}.
  
\bibitem{JonesWilsonDrainage}
  \Name{Jones A. F. \and Wilson S. D. R.}
  \REVIEW{J. Fluid Mech.}{87}{1978}{263}.       

\bibitem{JoyeAxisymmetrical}
  \Name{Joye J.-L., Miller C. A. \and Hirasaki G. J.}
  \REVIEW{Langmuir}{8}{1992}{3083}.
                                   
\bibitem{BertozziSingularities}
  \Name{Bertozzi A. L., Brenner M. P., Dupont T. F. \and Kadanoff L. P.}
  \Book{Trends and Perspectives in Applied Mathematics}  
  \Editor{L. Sirovich}
  \Publ{Springer-Verlag, New York}      
  \Year{1994}
  \Pages{155}{208}.
  
\bibitem{BergeronReview}
  \Name{Bergeron V.}
  \REVIEW{J. Phys.: Condens. Matter}{11}{1999}{R215}.  

\end{thebibliography}
\end{document}